\documentclass[%
reprint,
 amsmath,amssymb,
 aps,superscriptaddress,
pra,
]{revtex4-2}

\usepackage{spin-textures}

\usepackage{dcolumn}
\usepackage[T1]{fontenc}

\usepackage{booktabs} 

\usepackage[english]{babel}
\usepackage[utf8]{inputenc}
\usepackage[colorinlistoftodos]{todonotes}
\usepackage{chemformula} 

\DeclareSIUnit{\molar}{\textsc{M}} 

\newcommand{\CoNi}[2]{Ni\textsubscript{$#2$}Co\textsubscript{$#1$}}
\newcommand{\CoNiB}[2]{(Ni\textsubscript{$#2$}Co\textsubscript{$#1$})B}

\begin{document}
\preprint{APS/123-QED}

\title{Electrical characterization of the azimuthal anisotropy of \CoNiB{1-x}{x}-based ferromagnetic nanotubes}
\author{Dhananjay Tiwari}
\email{dtiwari899@gmail.com}
\affiliation{Univ. Grenoble Alpes, CNRS, CEA, SPINTEC, Grenoble, France}
\altaffiliation[Present address: ]{Advanced Safety and User Experience, Aptiv Services Poland SA, Krakow, Poland}
\affiliation{Univ. Grenoble Alpes, CEA, CNRS, Institut Néel, Grenoble, France}
\author{Martin Christoph Scheuerlein}
\affiliation{Department of Materials and Earth Sciences, Technical University of Darmstadt, Darmstadt, Germany}
\author{Mahdi Jaber}
\affiliation{Univ. Grenoble Alpes, CNRS, CEA, SPINTEC, Grenoble, France}
\author{Eric Gautier}
\affiliation{Univ. Grenoble Alpes, CNRS, CEA, SPINTEC, Grenoble, France}
\author{Laurent Vila}
\affiliation{Univ. Grenoble Alpes, CNRS, CEA, SPINTEC, Grenoble, France}
\author{Jean-Philippe Attané}
\affiliation{Univ. Grenoble Alpes, CNRS, CEA, SPINTEC, Grenoble, France}
\author{Michael Schöbitz}
\affiliation{Univ. Grenoble Alpes, CNRS, CEA, SPINTEC, Grenoble, France}
\affiliation{Univ. Grenoble Alpes, CEA, CNRS, Institut Néel, Grenoble, France}
\author{Aurelien Masseb\oe uf}
\affiliation{Univ. Grenoble Alpes, CNRS, CEA, SPINTEC, Grenoble, France}
\author{Tim Hellmann}
\affiliation{Surface Science Laboratory, Department of Materials and Earth Sciences, Technical University of Darmstadt, Darmstadt, Germany}
\author{Jan P. Hofmann}
\affiliation{Surface Science Laboratory, Department of Materials and Earth Sciences, Technical University of Darmstadt, Darmstadt, Germany}
\author{Wolfgang Ensinger}
\affiliation
{Department of Materials and Earth Sciences, Technical University of Darmstadt, Darmstadt, Germany}
\author{Olivier Fruchart}
\email{olivier.fruchart@cea.fr}
\affiliation{Univ. Grenoble Alpes, CNRS, CEA, SPINTEC, Grenoble, France}
\date{\today}

\begin{abstract}
We report on the structural, electric and magnetic properties of \CoNiB{1-x}{x} ferromagnetic nanotubes, displaying azimuthal magnetization. The tubes are fabricated using electroless plating in polycarbonate porous templates, with lengths several tens of micrometers, diameters from \SI{100}{\nano\meter} to \SI{500}{\nano\meter} and wall thicknesses from \SI{10}{\nano\meter} to \SI{80}{\nano\meter}. The resistivity is  $\sim\SI{1.5e-6}{\ohm\meter}$, and the anisotropic magnetoresistance~(AMR) of \SIrange{0.2}{0.3}{}\%, one order of magnitude larger~(resp. smaller) than in the bulk material, which we attribute to the resistance at grain boundaries. We determined the azimuthal anisotropy field from M(H) AMR loops of single tubes contacted electrically. Its magnitude is around \SI{10}{\milli\tesla}, and tends to increase with the tube wall thickness, as well as the  $\mathrm{Co}$ content. However, surprisingly it does not dependent much on the diameter nor on the curvature.
\end{abstract}


\maketitle

\section{Introduction}
\label{sec:Intro}
Nanotubes~(NTs) are hollow structures characterized by a sub-micrometer diameter, a wall thickness~(outer minus inner diameter), and a length much larger than the diameter. They are part of the wider family of one-dimensional structures (1-D), which in magnetism provide an ideal platform for both the fundamental investigation of domain-wall~(DW)\cite{bib-BOU2011} or skyrmion\cite{bib-FER2017} motion, spin-wave propagation\cite{bib-MAH2020}, and the implementation of logic\cite{bib-ALL2002,bib-KHI2010} or memory functionalities\cite{bib-PAR2008,bib-PAR2015b}. While most developments for magnetism in 1-D structure have been based on flat strips fabricated by the combination of physical deposition and nanofabrication so far, cylindrical structures offer specific physics related to curvature and dimensionality\cite{bib-FRU2018d}. For instance, a magnetic domain wall with a unique topology had been predicted to arise in nanowires and give rise to very high mobilities, the Bloch-point wall\cite{bib-HER2002a,bib-FOR2002,bib-THI2006}, whose existence and high mobility were recently confirmed experimentally\cite{bib-FRU2014,bib-FRU2019b}. NTs provide two additional degrees of freedom compared to nanowires, one being the ratio of outer over inner radius, the second being the ability to fabricate core-shell structures with interfaces. The latter is particularly appealing, as most spintronic effects arise from interfaces. One expects the magnetization to be uniform and parallel to the axis in long NTs made of a soft-magnetic material, because of the dipolar shape effect\cite{bib-LAN2009,bib-SUN2014}. Accordingly, theory predicted that the behavior of such magnetic NTs is very similar to that of magnetic nanowires, such as the occurrence of curling at the apex of the tube\cite{bib-ESC2007c}, and vortex-type domain walls with high mobilities\cite{bib-LAN2010,bib-OTA2013,bib-YAN2011b}. Other theoretical works examined the situation of tubes with azimuthal magnetization, predicting other specific features such as the curvature-induced non-reciprocal propagation of Daemon-Eschbach-type spin waves\cite{bib-OTA2017}.

From the experimental point of view there are now many methods for fabricating long NTs, based on the coating of porous anodized alumina\cite{bib-MAS2001} or polymer \cite{bib-MUeN2014} templates, or wire templates such as resulting from VLS growth\cite{bib-RUeF2014}. The coating methods include electrochemical deposition~\cite{bib-WAN2005b,bib-LI2008b}, atomic layer deposition (ALD)~\cite{bib-KNE2007}, electroless plating~\cite{bib-MUeN2014}, chemical vapour deposition (CVD)~\cite{bib-GOR2013b} or physical deposition. Yet another route for the fabrication of NTs is the nano-rolling of free thin films\cite{bib-STR2014}, however rather delivering diameters in the micrometer range. While the case of NTs with axial magnetization has been confirmed as expected\cite{bib-KIM2011b}, there have been a number of reports, demonstrating that domains with azimuthal magnetization could be obtained experimentally\cite{bib-RUeF2012,bib-STR2014,bib-FRU2018g,bib-ZIM2018,bib-POG2020}, either by coating non-magnetic wire templates by tilted-incidence physical deposition, rolled thin films or electroless plating of porous templates. While in the former two azimuthal magnetic anisotropy is reminiscent of the one arising in thin films induced by tilted deposition or uniaxial strain, the latter came more unexpectedly, and has been ascribed to the curvature-induced anisotropy of intergranular anisotropy or magneto-elastic energy\cite{bib-FRU2018g}. It is the purpose of the present work to report extensively on the link between azimuthal anisotropy in electroless-plated NTs versus tube diameter, wall thickness and material composition. The motivation is to provide a panorama of static properties that can be obtained, before searching for the magnetization dynamics predicted for NTs with azimuthal magnetization, and possibly  to shed light on the microscopic origin of magnetic anisotropy in such NTs.

\section{Synthesis and structural analysis}
\label{sec:synthesis}

\begin{figure}
\centering
\includegraphics[width=80mm]{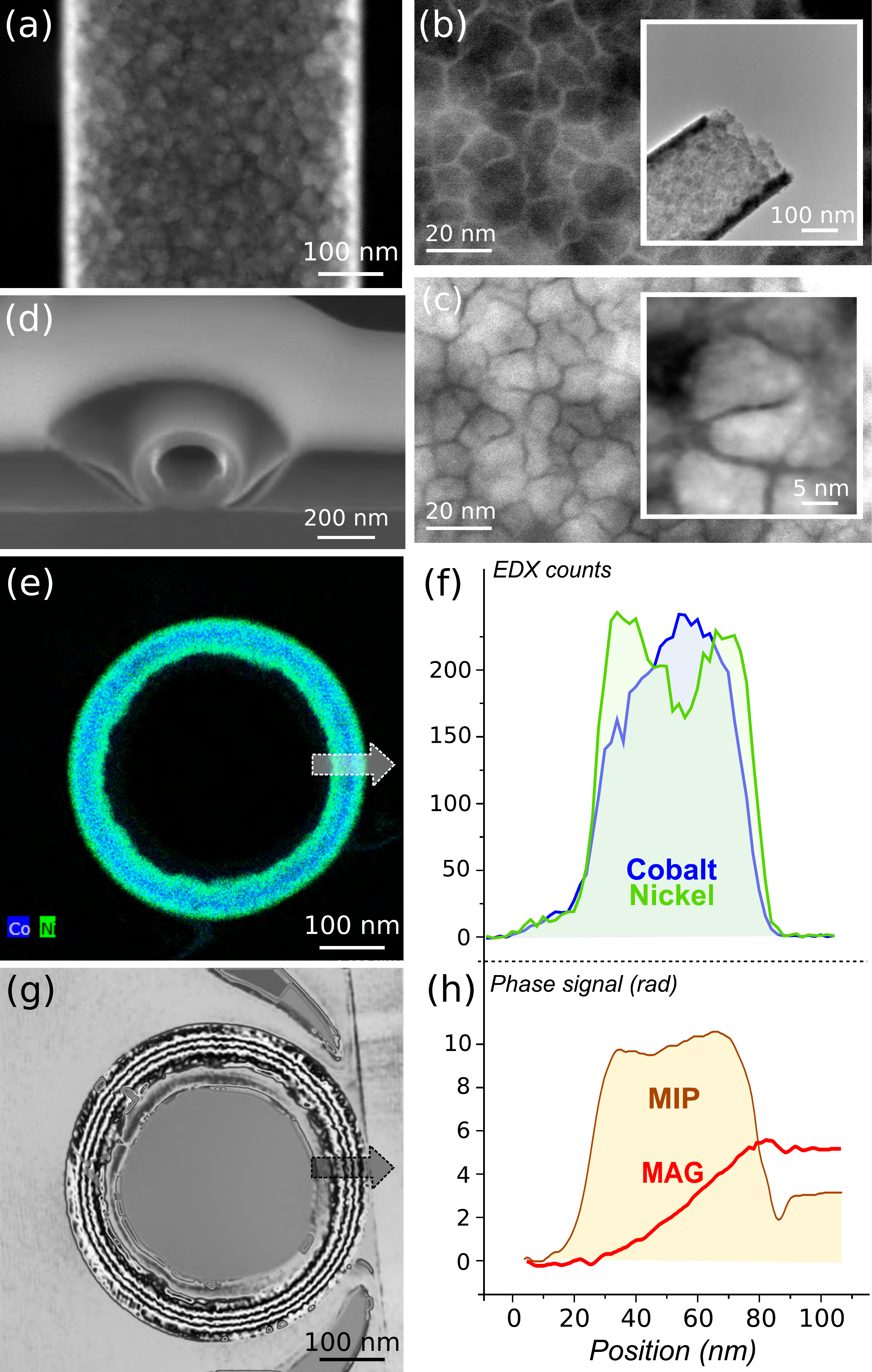}
\caption{\label{fig:Ni-Seg} TEM characterization of a NT. (a) STEM-HAADF view of a NT dispersed onto a grid, revealing a granular structure of the material. (b and c) Zoomed view in bright field and HAADF respectively of a one-wall-only end part, exhibiting the granular and intergranular structure spotted in the text. Insets display (c) the general view of the broken tube used for this single layer analysis and (d) higher magnification image in HAADF mode displaying the fine structure of the grains. (d) Penultimate step of the Focus Ion Beam preparation of a NT single slice. (e)~EDX mapping of the slice, highlighting the presence of Ni and Co (green and blue, respectively). The dashed arrow  indicates the extracted profiles. (f)~EDX profiles of the signal account for cobalt and nickel, using the same color code as for (c). (g)~Electron holography output, displaying $\mathrm{MIP} \times\cos \left( 5 \cdot  \mathrm{MAG}\right )$~(see text for details). Same dashed arrow as in (c), highlighting here the location for the phase profile. (h)~Phase profiles for the MIP (brown) and MAG (red) components of the phase shift~(see text for details).}
\end{figure}

Three batches of \CoNiB{1-x}{x} NTs ($x$ = 30, 50 and 80) were fabricated using electroless plating in ion track-etched polycarbonate membranes.
\begin{figure}
\centering
\includegraphics[width=85mm]{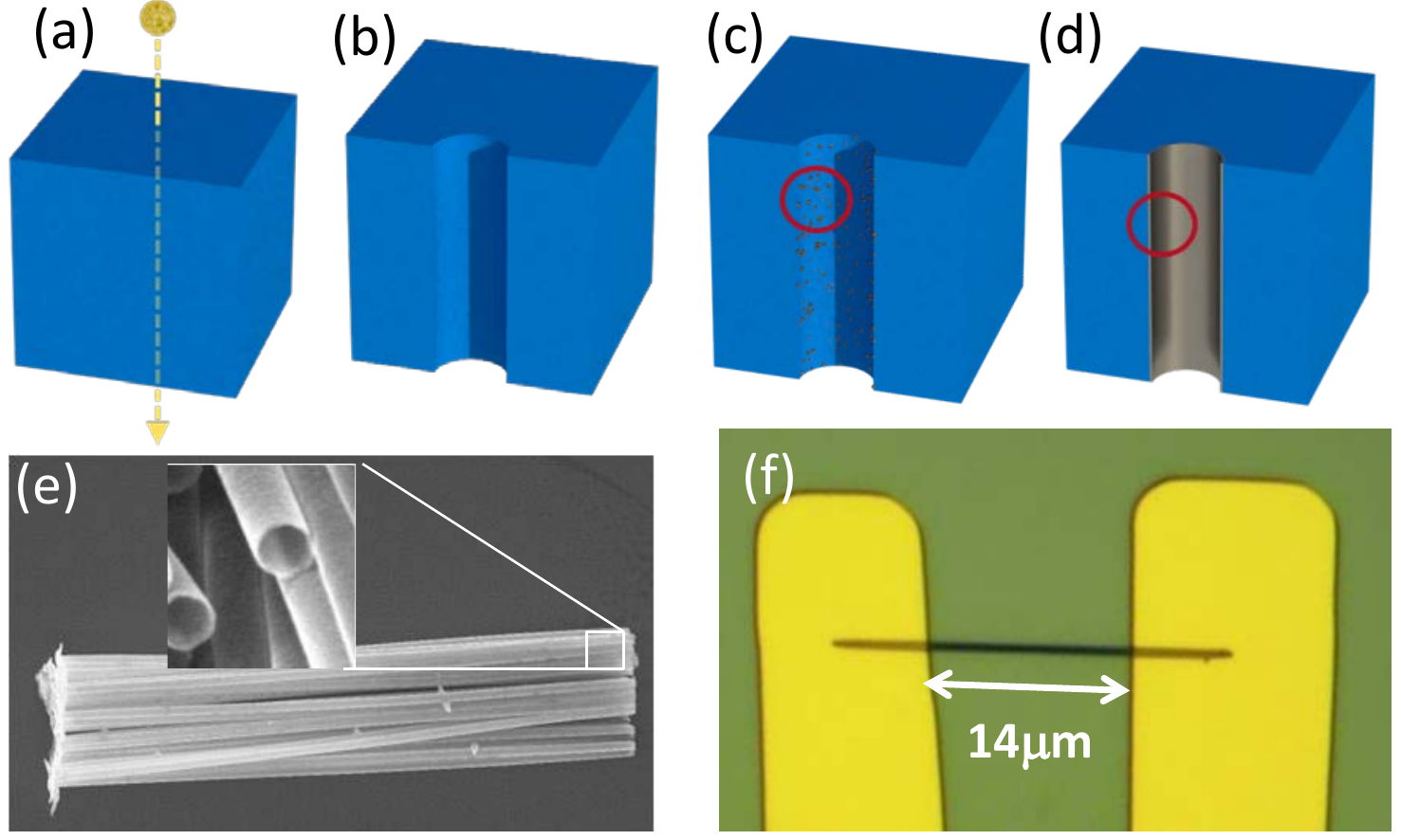}
\caption{\label{fig:SEM} Illustration of the NT growth process using electroless plating. (a) Damaged-track formation by swift heavy-ion irradiation. (b) Cylindrical pores formed inside polycarbonate membranes via chemical etching. (c) Formation of the active seed layer (Pd in our case) as a catalyst for material growth. (d) Reduction of metal ions for the growth of \CoNiB{1-x}{x} NTs, synthesized using electroless plating (e) Scanning electron microscopy (SEM) image of NTs released on a Si wafer from a drop of the solution with suspended NTs. (f) Optical image of a NT contacted electrically between two conductive pads.}
\end{figure}
The synthesis of the NTs is based on a previously-described procedure~\cite{bib-FRU2018g,bib-SCH2016c}, and is schematically shown in Fig.~\ref{fig:SEM} (a)-(d). First, polycarbonate foils are irradiated with swift heavy ions, creating latent damage tracks that are more vulnerable to chemical etching than the surrounding bulk polymer~[Fig.~\ref{fig:SEM}~(a)]. Subsequent treatment in a NaOH solution yields cylindrical pores, which are used as templates for the fabrication of NTs~[Fig.~\ref{fig:SEM}~(b)]. In order to initiate the electroless deposition reaction, catalytically-active Pd nanoparticules are deposited on the membrane surface by alternately submerging the membrane in Sn(II)- and Pd(II)-containing solutions~[Fig.~\ref{fig:SEM}~(c)]. Subsequently, the surfaces of the membrane are coated with \CoNiB{1-x}{x} by electroless plating, including the inside of the pores, yielding the formation of tubes. The Ni-to-Co ratio is tuned by changing the relative concentration of the respective metal precursors in the plating bath, while B is introduced a a byproduct by the reducing agent (dimethyl aminoborane, DMAB). A more detailed description of the NT fabrication and of the underlying mechanisms can be found in the appendix, section~\ref{Sec:deposition}. As indicated by XPS measurements, the B content of the material is in the range of \SI{20}{\percent~at}, which is typical for electroless CoB and NiB deposits fabricated using DMAB as a reducer (see Appendix, section~\ref{sec:XPS}).
After synthesis, the polycarbonate membranes were dissolved in dichloromethane. 
Ideally, this would yield a suspension of purely single, isolated NTs. However, since also the top and bottom surfaces of the membrane are coated during the electroless plating process, some of the tubes remain attached to one another~[see Fig.~\ref{fig:SEM}~(e)]. Nonetheless, due to the considerate amount of mechanical stress caused by the swelling of the polymeric matrix during dissolution, many single tubes are present in the suspension. Next, a drop of diluted NT suspension is applied onto highly-resistive silicon wafers, to obtain single NTs ready for further analysis, and ultimately electrical contacting, or to a copper grid with a lacey carbon film for TEM analysis.
Images obtained in conventional (S)TEM imaging are presented in Fig.~\ref{fig:Ni-Seg}~(a-c). The nano-granular structure of the grown layer is clearly visible. As transmission images of tubular structures overlap information from the top and bottom layer in the projected image, we focused on a broken tube [Fig.~\ref{fig:Ni-Seg}~(b-inset)] to conduct high-resolution imaging on a single layer. This delivers sharp images, from which the typical size of grains is inferred to be \SI{8 {\pm} 5}{nm} with a typical grain boundary as large as \SI{1}{nm}. The grain boundaries appear black in the HAADF contrast image [Fig.~\ref{fig:Ni-Seg}~(c)], which points at light elements, compatible with Boron (however, the detection of Boron was not possible at this high magnification in our setup). A higher magnification [Fig.~\ref{fig:Ni-Seg}~(c-inset)] image showed that the grains displays a finer structure, light with HAADF contrast, which we accordingly associate with the Pd seeds used for the electroless growth.\\

We used Focused Ion Beam to slice NTs and perform a cross-sectional analysis in a TEM. Fig.~\ref{fig:Ni-Seg}~(d) shows the TEM lamella before the final thinning, whose thickness we estimated to be around \SI{80 {\pm} 10}{nm} from know-how in such preparation. Observation of the thinned lamella indicates  a rather homogeneous wall thickness of \SI{60}{nm}. Energy-Dispersive X-ray (EDX) analysis of the tubes, which provides a qualitative yet not fully quantitative view, revealed that the $\mathrm{NiCo}$ composition is not homogeneous across the tube thickness. Instead, nickel tends to segregate towards both the inner and outer surfaces of the NT~[Fig.~\ref{fig:Ni-Seg}~(e) and (f)]. This surface enrichment in nickel comes with a decrease in cobalt content, an effect slightly more pronounced at the inner surface. We further analyzed the slice with Electron Energy Loss Spectroscopy (EELS). This revealed a variation of composition from \CoNi{50}{50} at the outer surface to \CoNi{60}{40} at the inner surface, separated by a plateau in the core of the material with a \CoNi{70}{30} composition. EELS also confirmed the absence of oxidation at the inner side of the NT\cite{jaber-unpub}. We also performed Electron Holography on the slice using the time-reversal~\cite{bib-DUN2004} method to separate the electrostatic (Mean inner potential - MIP) and the magnetostatic (MAG) parts of the reconstructed phase~[Fig.~\ref{fig:Ni-Seg}~(g) and (h)]. The iso-lines of the MAG-cosine phase are shown on Fig.~\ref{fig:Ni-Seg}~(g), displaying the magnetic induction flux lines. Fig.~\ref{fig:Ni-Seg}~(h) shows the MIP and MAG phases profiles. The slope of the latter is estimated to $\SI{0.11\pm0.04}{rad\per\nano\meter}$, which translates into an estimation of magnetization of $\muZero\Ms = \SI{0.9\pm0.1}{\tesla}$, based on slice thickness \SI{80}{nm}. Note that the MAG profile may indicate a slight decrease of magnetization (lower slope change) near the inner surface, consistent with the structural indication of lower Co content. However, this has not been seen uniformly on other profiles extracted at other parts of the slice. So, this could result from a local decrease of the thickness of the slice rather than from a composition change. Finally, it worth noticing that the flux-closure state observed in such slice cannot be extrapolated to a full tube as a ground state : the slicing strongly promotes azimuthal magnetization due to the short aspect ratio of the resulting tube, taking more the form of a ring here.

\section{Magnetic and transport properties}
\label{sec:discussions}
\subsection{Magnetotransport measurements}

To conduct transport measurement, NTs were transferred on highly-resistive silicon (Si) wafers ($\sim \SI{e6}{\ohm\meter}$), capped with natural oxide and pre-patterned with alignment marks.
The surface of the wafer is examined by scanning electron microscopy (SEM) to locate suitable NTs, with respect to the alignment marks. Next, one or a few  NTs per $\si{cm^2}$ were contacted electrically as follows. First, a two-lead pattern is written in the resist (positive resists, LOR 3A of $\sim\SI{200}{nm}$ and S1805 of $\sim\SI{500}{nm}$) using laser lithography.
Second, the surface of the NTs is cleaned through in-situ ion-beam etching, to remove any oxide layer from the surface. A Ti(\SI{15}{nm})/Au(\SI{250}{nm}) layer is then evaporated, followed by lift-off of the resist, which defines the conductive leads. The distance between two leads is \SI{14}{\micro\meter} in Fig.~\ref{fig:SEM}~(f). Details of the contacting process were already provided elsewhere~\cite{bib-FRU2019b}.

\begin{figure}
\centering
\includegraphics[width=85mm]{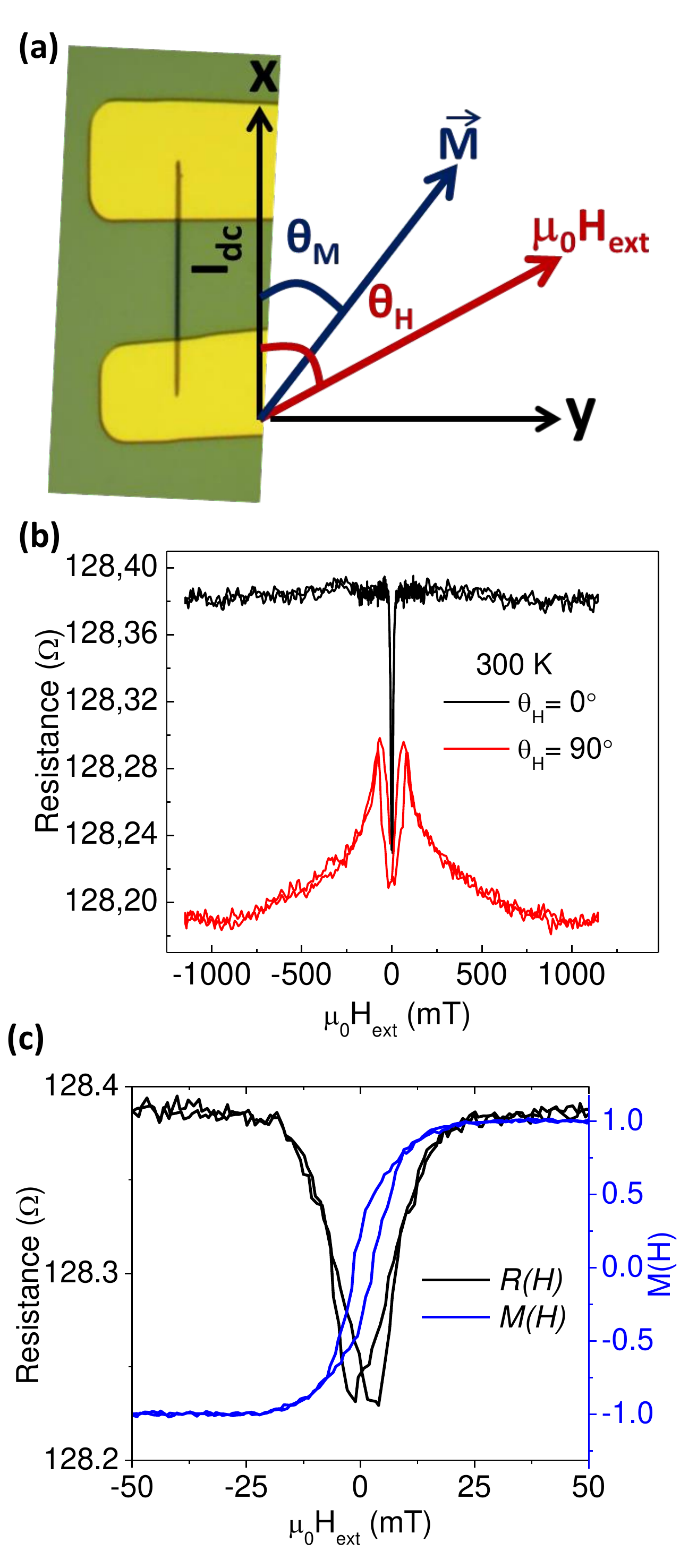}
\caption{\label{fig:set-up} (a)~Optical image of a contacted NT with composition \CoNiB{70}{30}, external diameter $d=\SI{470}{\nano\meter}$, wall thickness $t_0=\SI{59}{\nano\meter}$, and distance between the leads \SI{14}{\micro\meter}. (b)~Resistance versus applied field $\muZero\Hext$ swept up-and-down in a four-quadrant fashion, applied either parallel~(black) and perpendicular~(red) to the NT axis. (c) Hysteresis loop of \CoNiB{70}{30} tubes, displaying the longitudinal magnetization $M(H)=\cos[\theta_M(H)]$ reconstructed from $R(H)$ loops. \ie,  azimuthal magnetization.}
\end{figure}

Fig.~\ref{fig:set-up} shows the geometry of the measuring setup, and the magnetotransport characterization of a NT with composition \CoNi{70}{30}, tube diameter $d=\SI{470}{\nano\meter}$ and wall thickness $t_0=\SI{59}{\nano\meter}$. Transport measurements were conducted by applying a current ($I_{\rm DC}=\SI{10}{\micro\ampere}$), and measuring the voltage across the same leads. This corresponds to a current density of about \SI{1.3e8}{\ampere\per\meter\squared}, if assumed to be uniform across the tube.

The resistance of the device at remanence is \SI{128.3}{\ohm} at \SI{300}{K} and \SI{112.3}{\ohm} at \SI{10}{K}. This translates into resistivities $\rho_0=\SI{1.5e-6}{\ohm\meter}$ at \tempK{300} and  $\rho_0=\SI{1.3e-6}{\ohm\meter}$ at \tempK{10}, assuming absence of contact resistance and of voltage drop across the leads. These values are one order of magnitude higher compared to bulk NiCo\cite{bib-McGU1975}. This likely results from the high boron content and from the nanocristalline nature of the material~[Fig.~\ref{fig:Ni-Seg}~(a)], liable to give rise to inter-granular resistance at grain boundaries. This will be further supported by magnetoresistive measurements, reported below.

Magnetoresistance properties were investigated by applying an external magnetic field up to $\SI{1}{T}$ along various directions. The geometry is sketched in Fig.~\ref{fig:set-up}~(a), with $\theta_H$~(resp. $\theta_M$) the angle between the applied field~(resp. magnetization) and the axis of the NT, which is also the direction of the flowing current. The measurements~[Fig.~\ref{fig:set-up}~(b)] are qualitatively similar to those already performed on various types of NTs displaying azimuthal magnetization\cite{bib-RUeF2012,bib-RUeF2014}, which we analyze in the following. For magnetic field applied along the tube axis~[black in Fig.~\ref{fig:set-up}~(b), with zoom in Fig.~\ref{fig:set-up}~(c)], saturation is reached at about \SI{20}{\milli\tesla}. The $R(H)$ loop is qualitatively consistent with the picture of azimuthal magnetization at remanence already proven directly in the same tubes by magnetic imaging\cite{bib-FRU2018g}, and with the existence of a positive anisotropic magnetoresistance in these materials~(AMR)\cite{bib-McGU1975,bib-OHA2000}
as defined by:
\begin{eqnarray}
  \label{e1} R = R_{\perp} + (R_{\parallel} - R_{\perp}) \cos^{2}\theta_{\rm M} ,\\
  \label{e2} \mathrm{AMR} = \frac{R_{\parallel}-R_{\perp}}{R_{\perp}} .
\end{eqnarray}
The magnetoresistance curve is slightly hysteretic at small fields, meaning that the direction of magnetization depends on the magnetic history. This implies that magnetization may not be perfectly azimuthal at remanence. To illustrate the rotation of the magnetization under application of the longitudinal field, is is convenient to display normalized $M(H)$ loops obtained as $\cos\theta_{\rm M} = \sqrt{\Delta R(H)}/\Delta R_{\rm AMR}$, with $\Delta R(H)=R(H)-R_0$ and $\Delta R_{\rm AMR}=R_{\rm sat}-R_0$, with the resistances $R_{\rm sat}$ at saturation, and $R_0$ the minimum resistance~[Fig.~\ref{fig:set-up}~(c)]. The hysteresis now appears in a more usual fashion, with coercivity of about~\SI{2}{\milli\tesla}.

\begin{figure}
\centering
\includegraphics[width=42.712mm]{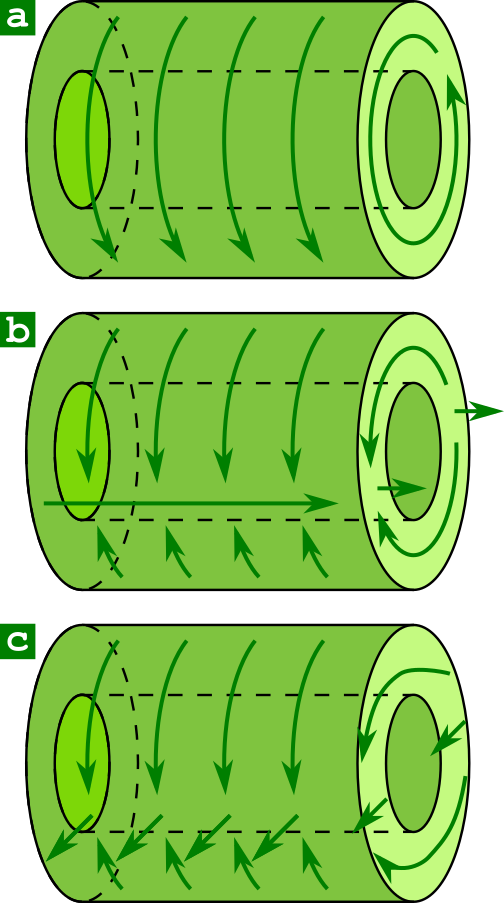}
\caption{\label{fig:magn-states} Sketch of the expect magnetization state under transverse applied magnetic field: at (a)~remanence, fully azimuthal (b)~at intermediate field, in an onion state (c)~at large field with asymptotically uniform magnetization.}
\end{figure}

We now examine a magnetoresistance curve for a magnetic field applied across the tube, \ie, $\theta_H = 90^{\circ}$~[red line in Fig.~\ref{fig:set-up}~(b)]. Starting from remanence the resistance increases until $\muZero\Hext \approx\SI{65}{mT}$, and then continuously decreases up to $\SI{1}{T}$. This bell-shaped response shares some features with previously-reported magnetoresistive curves of NTs with azimuthal magnetization\cite{bib-RUeF2012,bib-RUeF2014}, however here with a much clearer dip at remanence. We understand the bell shape the following way~(Fig.\ref{fig:magn-states}). At remanence the value of resistance is very similar to that obtained with a longitudinal field, which points at sharing the same remanent state, with a largely azimuthal magnetization. When the transverse field is increased a transition to an onion state is expected\cite{bib-RUeF2012}. This process is likely to explain the sizable hysteresis around this field. Indeed, the change of magnetization distribution cannot be achieved reversibly, and requires nucleation and motion of domain walls. At the transition field the head-to-head and tail-to-tail parts are not expected to be aligned along the applied field, which is moderate, but rather rotate along the axial direction to remain parallel to the local surfaces, and thereby keep the magnetostatic energy moderate. Magnetization in these parts is expected to be parallel to the electric current, which is consistent with the increase of resistance. This area with axial magnetization is expected to decrease upon increasing the field, ending in a NT mostly saturated transverse to its axis. Indeed, in this situation magnetization is perpendicular to the current everywhere, bringing resistance to a minimum. It is the difference between the maximum and minimum values on Fig.~\ref{fig:set-up}~(b), considering all directions of applied field, that defines most accurately the magnitude of AMR, based on \eqnref{e2}. This sets the AMR ratio of the single \CoNiB{70}{30} NT at $\sim\SI{0.15}{\%}$ at \tempK{300} and $\SI{0.25}{\%}$ at \tempK{10}. These figures are one order of magnitude lower than both bulk CoNi and CoNiB alloys\cite{bib-McGU1975}.
This is consistent with the high resistivity measured, understood as a resistance dominated by intergranular effects. Indeed, the latter should not give rise to magnetoresistance as long as magnetization is uniform along the current flow, which is largely expected here.

The same measurements have been made on individual NTs with concentration \CoNiB{50}{50} and \CoNiB{80}{20}. The behavior is qualitatively similar, with the quantitative analysis reported in the next section.

\subsection{Magnetic anisotropy}
\begin{figure}
\centering
\includegraphics[width=85mm]{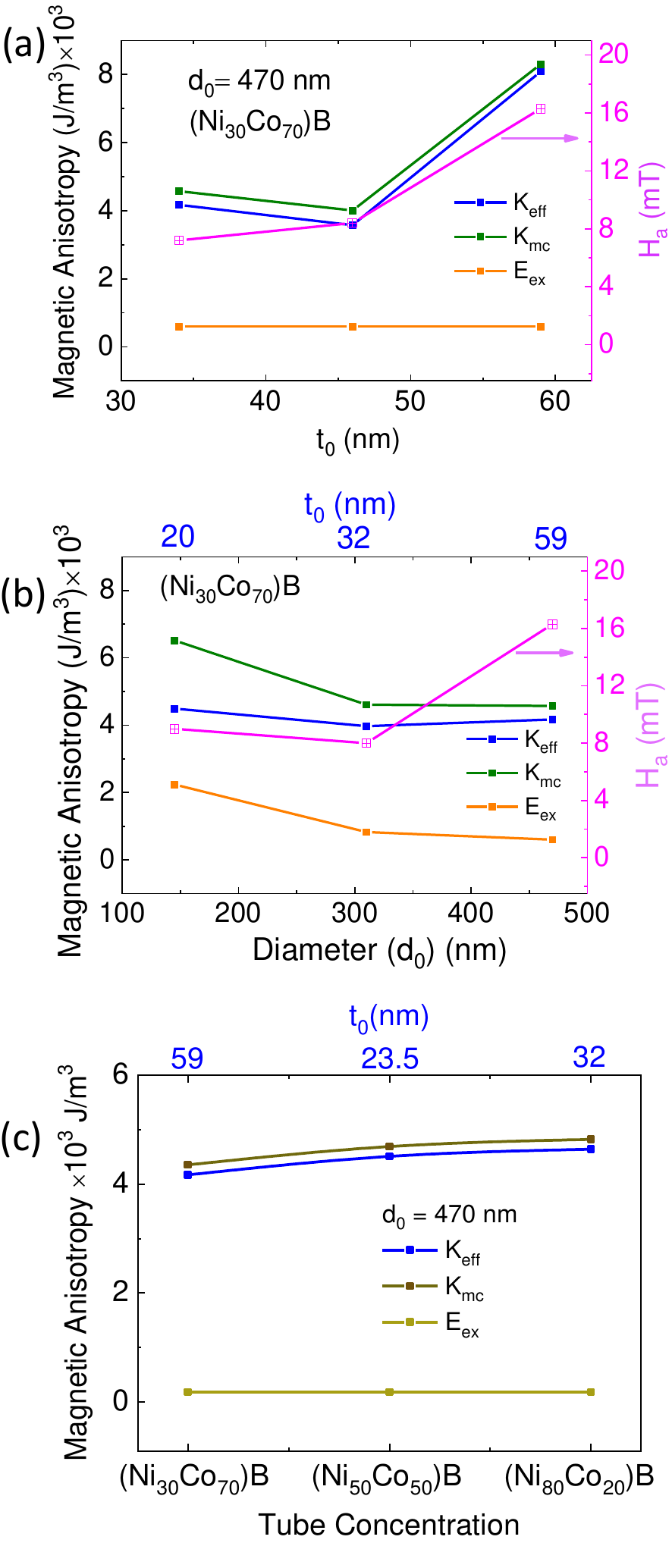}
\caption{\label{fig:anisotropies} Magnetic anisotropies:
effective anisotropy volume density $K_\mathrm{eff}$ and anisotropy field $H_{\mathrm a}$ determined experimentally, versus: (a)~tube thickness, (b)~tube diameter, and (c)~material composition. The measured tube thickness measured for each sample is indicated on the top $x$ axis for (b) and (c)~(see text). $H_\mathrm{a}$ represents the extracted magnetic anisotropy field.}
\end{figure}
We now report on the determination of the strength of the azimuthal magnetic anisotropy of the NTs derived from their hysteresis loops as reconstructed in Fig.~\ref{fig:set-up}~(c), and then discuss it versus the geometry and composition of the NTs.

We first describe the protocol to determine the various quantities associated with the anisotropy. Owing to the moderate thickness of the NTs considered in the following, we assume that the direction of the magnetization may not vary significantly across the radius, and accordingly consider an effective value of volume density of magnetic anisotropy, $K_{\rm eff}$. The volume density of magnetic energy of a NTs is $K_{\rm eff}\cos^2\theta_M$, $\theta_M$ being the polar angle of magnetization versus the tube axis. With these notation, a positive value for $K_{\rm eff}$ means a longitudinal hard axis. Considering the local shape anisotropy that shall tend to restrict the magnetization direction within the shell, a positive $K_{\rm eff}$ translates into an azimuthal easy axis. This effective anisotropy is, by definition, the area above the magnetization hysteresis loop considered along a hard-axis direction:
\begin{equation}\label{eqn-anisotropyFromLoops}
    K_{\rm eff}=\muZero\int_0^{\Ms} H(M)\,\mathrm{d}M.
\end{equation}
To avoid a calculation bias induced by the hysteresis, even if moderate, we consider the unhysteretic curve by averaging the up and down H(M) curves.

The strength of the effective magnetic anisotropy may also be expressed in terms of the anisotropy field, defined as:
\begin{equation}\label{e3}
    H_{\rm a}=\frac{2K_{\rm eff}}{\mu_{\rm 0}M_{\rm s}}.
\end{equation}
Calculating these values requires information about the magnetization of the material properties. This was done using magnetometry, with good agreement with the Slater-Pauling curve for CoNi alloys~(see appendix).

Finally, for the sake of identifying the role of curvature in the anisotropy, it is important to remember that we expect two contributions to $K_{\rm eff}$. The first contribution arises from the interaction with the lattice, which we will write $K_{\rm mc}$ for magnetocrystalline anisotropy, be it a magnetocrystalline, magnetoelastic or interface anisotropy. The second contribution is that of the exchange energy ($K_{\rm ex}$) associated with azimuthal curling of the magnetization:
\begin{equation}\label{eqn-exchangeForCurling}
    K_{\rm ex}=-\frac{A}{R_0^2}\;,
\end{equation}
with $R_0$ the average tube radius\cite{bib-FRU2018d,bib-SUN2014,bib-SAX1998} and $A\approx\SI{10}{\pico\joule\per\meter}$ the exchange stiffness. The minus sign reflects the fact that exchange favors axial uniform magnetization as the ground state. So, in the end the anisotropy arising from the lattice, \ie, the microstructure of the material, is $K_{\rm mc}=K_{\rm eff}-K_{\rm ex}$.

We now present and discuss the values of the azimuthal anisotropy. The order of magnitude of $K_\mathrm{eff}$ and $\muZero\Ha$ are $\SI{5}{\kilo\joule\per\cubic\meter}$ and \SI{10}{\milli\tesla}, respectively. Fig.~\ref{fig:anisotropies} display the dependence of the volume density of azimuthal magnetic anisotropy, as well as that of the associated anisotropy field, versus the tube thickness, diameter and material composition. Note that at the synthesis stage we aim at reaching a nominal thickness by controlling the deposition time, however in practice the thickness may deviate from the target and needs be determined by TEM. This explains why we cannot display the diameter and composition dependence fully independently from the thickness, which is indicated at the top $x$ axis for (b) and (c), for each sample. Fig.~\ref{fig:anisotropies}(a) displays the anisotropy versus tube thickness dependence for fixed diameter $d_0=\SI{470}{nm}$ and composition \CoNiB{70}{30}, for different deposition time. The contribution of the exchange to $K_\mathrm{eff}$ is negligible for this large diameter. The anisotropy tends to increase for larger thicknesses. This is understandable as strain, curvature and therefore the anisotropy of grains are expected to increase as the inner diameter of the tube decreases. More surprising is the presence of a plateau between 35 and \SI{45}{nm}. Fig.~\ref{fig:anisotropies}(b) displays the anisotropy versus the tube diameter, for a thickness of about \SI{30}{nm} and composition \CoNiB{70}{30}. The contribution of exchange is weak, except for the smaller diameters investigated here, \ie, \SI{150}{nm}. Following the subtraction of the contribution of exchange, the part of anisotropy due to the lattice only, $K_\mathrm{mc}$, does not show a clear variation with the curvature. This is surprising as curvature is a required ingredient to break the symmetry between the axial and azimuthal directions, and therefore induce azimuthal anisotropy. Last, Fig.~\ref{fig:anisotropies}(c) displays the anisotropy versus composition for a diameter \SI{470}{nm} and thickness of about \SI{30}{nm}. Again, the contribution of exchange is negligible for this large diameter. As one possible underlying physical mechanism for anisotropy is strain and inverse magnetostriction, let us examine what is known about CoNi alloys. Both magnetostriction coefficients $\lambda_{100}$ and $\lambda_{111}$ of CoNi metallic single crystals increase with the Co concentration in the present range\cite{bib-KAD1981}. Regarding boron-containing alloys, data is available for metallic glasses\cite{bib-OHA1978}, showing a maximum of magnetostriction around the composition \CoNi{50}{50}. Both situations would be consistent with the smaller anisotropy of Ni-rich alloys, although one should remain cautious about the interpretation, as the fine details of the electroless material are not known~(eg., the exact amount of boron, and whether in the matrix or at the grain boundaries). Also, the surface segregation of nickel and the enrichment with Co~(Fig.\ref{fig:Ni-Seg}) of the core may also affect anisotropy.

\section{Conclusion}
We have investigated the magnetoresistive properties and the strength of the magnetic anisotropy favoring the azimuthal direction of magnetization in electroless-plated  \CoNiB{1-x}{x} nanotubes, versus the nanotube diameter and thickness, and the composition of the alloy. The measured resistivity is one order of magnitude  higher compared to that of bulk samples, while the anisotropic magnetoresistance is one order of magnitude lower, which we believe is related to the drop of voltage across grain boundaries in this nanocrystalline material. The strength of the azimuthal magnetic anisotropy is of about \SI{5}{kJ\per\cubic\meter}. The anisotropy tends to increase with the tube thickness, and depends only weakly on its diameter. While no direct proof can be given about its microscopic origin, its variation with the material composition is consistent with the curvature-induced anisotropy of strain, combined with inverse magnetostriction.
\acknowledgments
This project received support from the ANR-DFG C3DS project (ANR-18-CE92-0045, DFG-406700532). A CC-BY public copyright license has been applied by the authors to the present document and will be applied to all subsequent versions up to the Author Accepted Manuscript arising from this submission, in accordance with the grant’s open access conditions\cite{bib-CC-BY}. We thank the Nanofab platform at Institut Néel, whose team and especially Bruno Fernandez provided technical support. M. C. S. and T. H. are grateful for valuable discussion with Jona Schuch (Surface Science Laboratory, TU Darmstadt). M. C. S. and W. E. thank Prof. Christina Trautmann and Dr. Maria Eugenia Toimil-Molares (Materials Research Group, GSI Helmholtzzentrum für Schwerionenforschung) for their support during the ion irradiation experiments.
\appendix
\section{Resistivity measurements}
\label{Sec:resistivity}
\begin{figure}[h!]
\centering
\includegraphics[width=85mm]{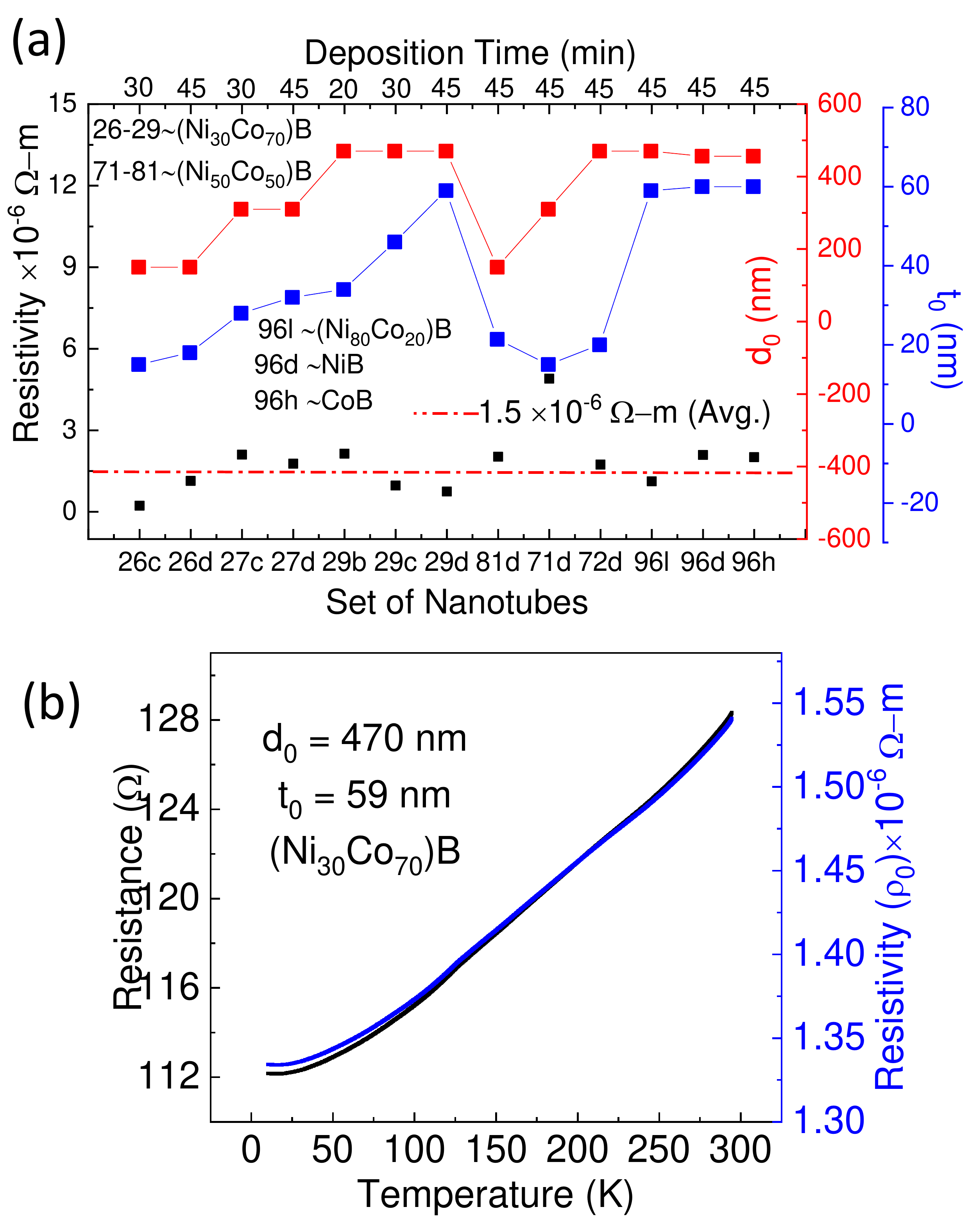}
\caption{\label{fig:Resistivity} (a) Electrical resistivity of NTs versus their geometrical features and chemical compositions. (b) Resistivity as a function of temperature.}
\end{figure}

The evaluation of the resistivity is based on the measured resistance, and requires knowledge of the cross-sectional area of the NT through which the current is flowing. Thus, uncertainties on both affect the inferred value for resistivity. The value of resistance may be affected by a number of effects, notably an interfacial resistance between the tube and the electrical leads. As regards area, in electroless plating, the tube thickness ($t_{\rm 0}$) depends on the duration of deposition time, on the concentration of the solution, and the diameter of the pore, the latter hindering diffusion. Thus it is important to measure directly the thickness of every batch, and in practice this provides some variety of values for the thickness\cite{bib-FRU2018g,bib-SCH2016c}. Figure~\ref{fig:Resistivity}~(a) provides an overview of the resistivity measurements on NTs with various thicknesses, diameters and composition, showing a spread in results, however not clearly correlated with any of these parameters. Fig.~\ref{fig:Resistivity}~(b) shows that resistivity decreases at low temperature, consistent with a metallic behavior. However, the decrease is moderate compared with a clean metal. This confirms our hypothesis of resistance dominated by inter-granular resistance due to boron-rich grain boundaries, expected to be only weakly dependent on temperature. This phenomenon is phenomenologically similar to the usual situation of grain boundary scattering in metals, which induces an offset in resistivity, however without affecting much its temperature dependence\cite{bib-MAE1965,bib-VRI1988}.

\section{Magnetization measurements}
\label{Sec:Magnetization_tube_concentration}
\begin{figure}[h!]
\centering
\includegraphics[width=85mm]{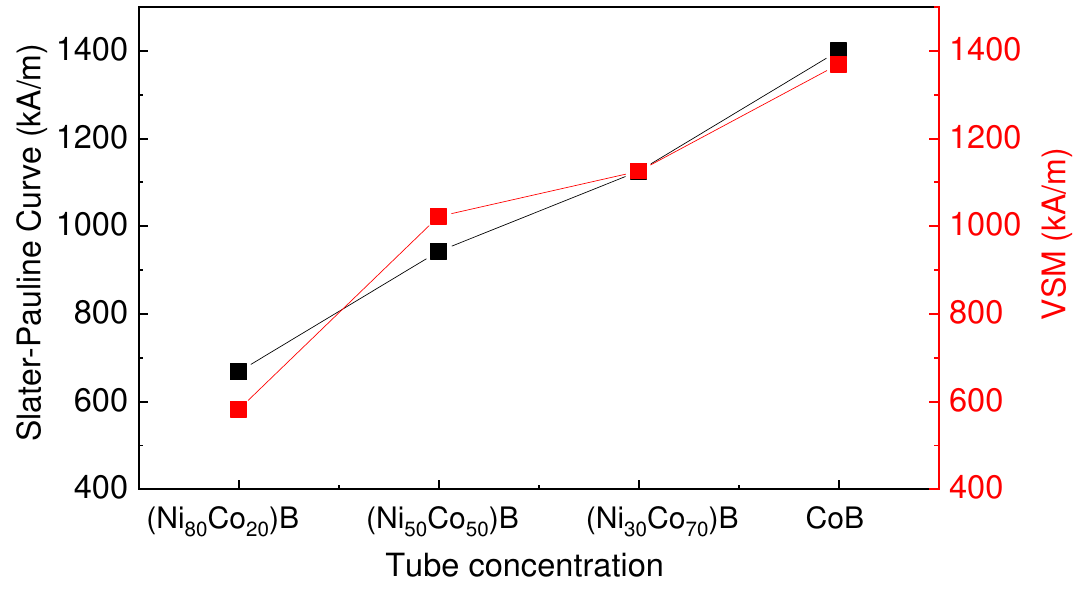}
\caption{\label{fig:Magnetization_tube_concentration} The experimental values of $M_{\rm s}$ as a function of tube concentration \CoNiB{1-x}{x}}
\end{figure}

Magnetization has been inferred for all compositions reported here based on hysteresis loops of thin films of a given area, measured by vibrating sample magnetometry~(Fig.~\ref{fig:Magnetization_tube_concentration}). The variation is very similar to that expected from the linear variation of the Slater-Pauling curve~\cite{bib-CUL2011} for pure CoNi alloys, \ie, with no boron. This hints at a rather low boron concentration in the material.
\section{Anisotropy field confirmed from magnetometry}
Hysteresis loops were performed with vibrating sample magnetometry on an array of \CoNiB{70}{30} tubes still in the  polycarbonate membrane, with magnetic field applied along the tube axis~(\ie, perpendicular to the polymer foil). Compared with the situation of single tubes investigated via AMR, magnetizing all tubes along their axis requires to pay the cost of the demagnetizing energy of the entire array of tubes, scaling with the magnetic filling factor. This translates into a contribution to the anisotropy energy :
\begin{equation}\label{eqn-magneticFillingFactor}
 H_{\rm a,tube-array} = M_{\rm s}\times \rho_{\rm d}\times\pi(R_0^2-R_\mathrm{i}^2)\;.
\end{equation}
$\rho_{\rm d}=\SI{e8}{\per\square\centi\meter}$ is the areal density of pores, $\pi(R_0^2-R_\mathrm{i}^2)$ the cross-sectional area of the tube with parameters $R_{\rm 0} = \SI{235}{nm}$ the outer radius and $R_\mathrm{i}=\SI{205}{nm}$ the inner radius. The calculated $H_{\rm a,tube-array}$ using Eq(C1) is 207 mT and  $H_{\rm a,VSM}$ = 186 mT. The magnitude of anisotropy field of a single tube is then derived as: $H_{\rm a,single-tube}=H_{\rm a,VSM}-H_{\rm a,tube-array}$. Its numerical value is $\sim\SI{20}{mT}$, which is quantitatively similar to $H_{\rm a}=\SI{16.3}{mT}$ measured from the reconstructed M-H loop as shown in Fig.~\ref{fig:set-up}~(c). The latter is however more reliable, not requiring a subtraction between too large figures when tubes are interacting in the array.
\section{Effect of annealing}
\label{sec:Annealing}
\begin{figure}[h!]
\centering
\includegraphics[width=85mm]{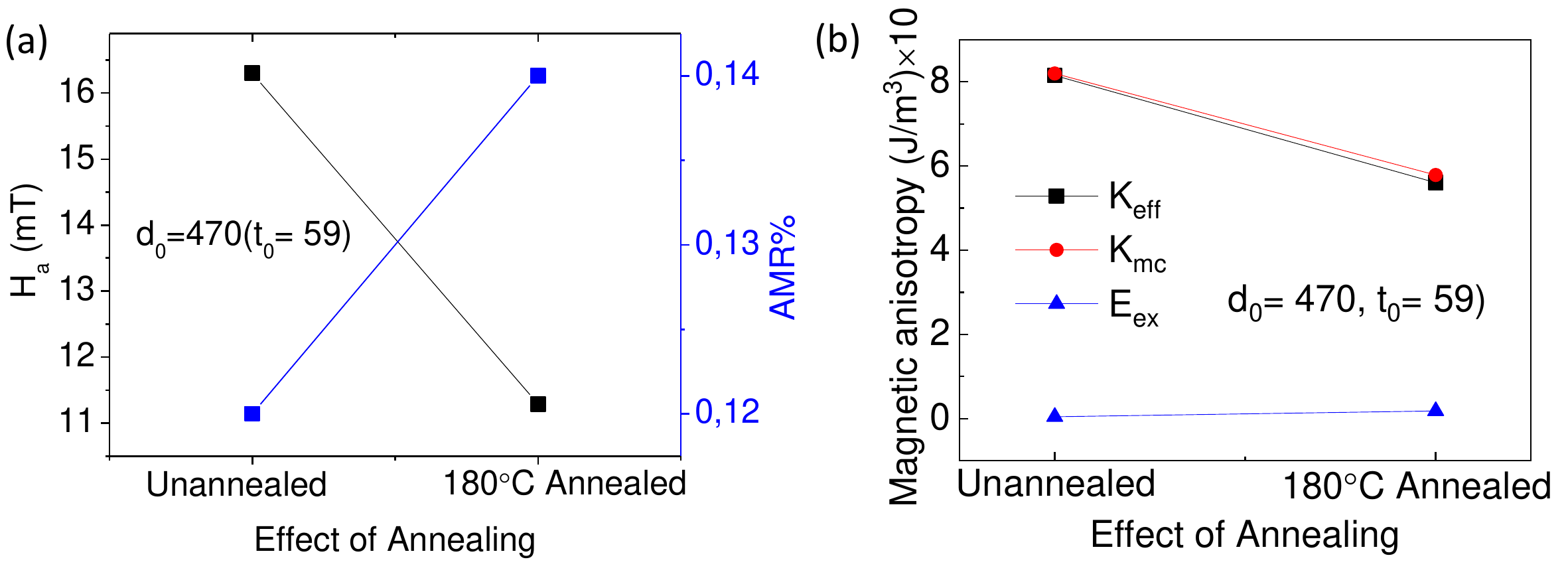}
\caption{\label{fig:Annealing} Effect of annealing on anisotropies for $(\textrm{Ni}_{\rm 30}\textrm{Co}_{\rm 70})\textrm{B}$ tubes. (a) $H_{\rm a}$ and AMR (b) magnetic anisotropies as a function of annealing.}
\end{figure}
It was previously shown that annealing tends to decrease the strength of magnetic anisotropy in these NTs, ultimately restoring axial magnetization\cite{bib-FRU2018g}. Here, a batch of NTs was annealed at $180^{\circ}$ for 2 hours before contacting the tubes. We observe that $H_{\rm a}$ and $K_{\rm eff}$ decrease with annealing, as expected~[Fig.~\ref{fig:Annealing}]. This is observed for all concentration of $(\mathrm{Ni}_{x}\mathrm{Co}_{1-x})\mathrm{B}$ NTs. Also, the resistivity decreases, which is consistent with grain growth and our hypothesis that resistance is dominated by grain boundaries.
\section{Electroless deposition of \ch{(Ni_{x}Co_{1-x})B} nanotubes}
\label{Sec:deposition}

The synthesis of \ch{(Ni_{x}Co_{1-x})B} NTs was performed according to a previously published procedure, with minor modifications~\cite{bib-FRU2018g,bib-SCH2016c}. The sample fabrication is briefly described in the following paragraphs.

\subsection*{Chemicals and methods}

The following chemicals have been used, without further modification or purification: Tin(II) chloride dihydrate (\textit{Sigma-Aldrich}, \SI{98}{\percent}), trifluoroacetic acid (\textit{Sigma-Aldrich}, \SI{99}{\percent}), methanol (\textit{PanReac AppliChem}, pure), palladium(II) chloride (\textit{Aldrich}, \SI{99}{\percent}), potassium chloride (\textit{PanReac Applichem}, USP, Ph. Eur.), nickel(II) sulfate heptahydrate (\textit{Acros Organics}, for analysis), cobalt(II) sulfate heptahydrate (\textit{Sigma-Aldrich}, $\ge$\SI{98}{\percent}), trisodium citrate (\textit{Alfa-Aesar}, \SI{99}{\percent}), borane dimethylamine complex (DMAB, \textit{Aldrich}, \SI{97}{\percent}).\\

All aqueous solutions were prepared using purified water (\textit{Milli-Q}, $>$\SI{18.2}{\mega\ohm}). Prior to use, all glassware was cleaned with boiling \textit{aqua regia}, stored in an alkaline bath for multiple days and rinsed with copious amounts of deionized water.

\subsection*{Template preparation}

Polycarbonate (PC) foils with a thickness of \SI{30}{\micro\meter} (Pokalon, \textit{Lofo High Tech Film GmbH}) were irradiated with swift heavy ions (\ch{Au^{26+}}, \SI{5.9}{\MeV\per u}, \SI{1d8}{\centi\meter\tothe{-2}}) using the \textit{UNILAC} linear accelerator facility at \textit{GSI Helmholtzzentrum für Schwerionenforschung GmbH}, Darmstadt, Germany~[Fig.~\ref{fig:SEM}(a)]. Cylindrical pores were obtained by subsequent chemical etching in stirred, aqueous \SI{6}{\molar} NaOH solution at $\SI{50}{\degreeCelsius}$~[(]Fig.~\ref{fig:SEM}(b)]. The duration of the etching process, which determines the diameter of the pores, was varied between \SIrange{10}{30}{\minute}.

\subsection*{Electroless deposition}

To initiate the electroless plating reaction, catalytically-active Pd seeds are deposited on the PC template surface by a previously described, two-step sensitization and activation procedure:~\cite{bib-SCH2016c} Firstly, the membranes are submerged in a Sn(II)-containing solution [\SI{42}{\milli\molar} \ch{SnCl_2 * 2 H_2O} and \SI{72}{\milli\molar} trifluoroacetic acid in methanol and water (1:1)] for \SI{45}{\minute}. After washing with water, they are transferred to an aqueous Pd(II)-solution (\SI{11.3}{\milli\molar} \ch{PdCl_2}, \SI{33.9}{\milli\molar} \ch{KCl}) for 4 minutes. The two steps are repeated two more times, with the sensitization duration shortened to \SI{15}{\minute}. Electroless plating was then conducted from a bath containing \ch{NiSO_4 * 7 H_2O} and \ch{CoSO_4 * 7 H_2O} as the metal-ion source (\SI{100}{\milli\molar} in total), disodium citrate as a chelating ligand (\SI{100}{\milli\molar}), as well as borane dimethylamine (DMAB) as reducer (\SI{100}{\milli\molar}). The Ni/Co ratio of the final \ch{(Ni_{x}Co_{1-x})B} deposit was determined by the ratio of the respective metal-ions in the plating solution, while the wall-thickness of the tubes was controlled by the deposition time. Due to the faster plating speed of Ni-rich electrolytes (see Appendix~\ref{Sec:deposition}) the depositions for \ch{(Ni_{0.5}Co_{0.5})B} and \ch{(Ni_{0.8}Co_{0.2})B} were conducted at \SI{4}{\degreeCelsius}, all others at room temperature ($\sim\SI{25}{\degreeCelsius}$).

\subsection*{Tuning the composition of \ch{(Ni_{x}Co_{1-x})B} nanotubes}

Due to the similar chemical behavior of \ch{Co^2+} and \ch{Ni^{2+}} ions, it is possible to deposit alloys of the respective metals from a single plating bath. In both cases, citrate has proven to be a suitable ligand and stabilizer, and both metals are catalytically active towards DMAB decomposition. This allows to tune the Ni/Co ratio of the deposit by simply adjusting the ratio of ions in solution. Comparing the deposition reactions of Ni-rich and Co-rich electrolytes with the same reactant concentrations, it can be observed that the deposition of the Ni-rich materials is considerably faster than that of their Co-rich counterparts. This can be attributed to the slightly more positive reduction potential of Ni,~(see Equation~\ref{eq:reductionNi} and~\ref{eq:reductionCo})~\cite{bib-HAY2014}, as well as its higher catalytic activity towards DMAB decomposition~\cite{bib-SAI1998,bib-OHN1985}.

\begin{align}
    \label{eq:reductionNi}
  \ch{Ni^{2+} + 2 e^- -> Ni}; & & \SI{-0.257}{\volt}~\mathrm{vs.~SHE}\\
  \label{eq:reductionCo}
  \ch{Co^{2+} + 2 e^- -> Co}; & & \SI{-0.28}{\volt}~\mathrm{vs.~SHE}
\end{align}

\begin{figure}[h!]
\centering
\includegraphics[width=85mm]{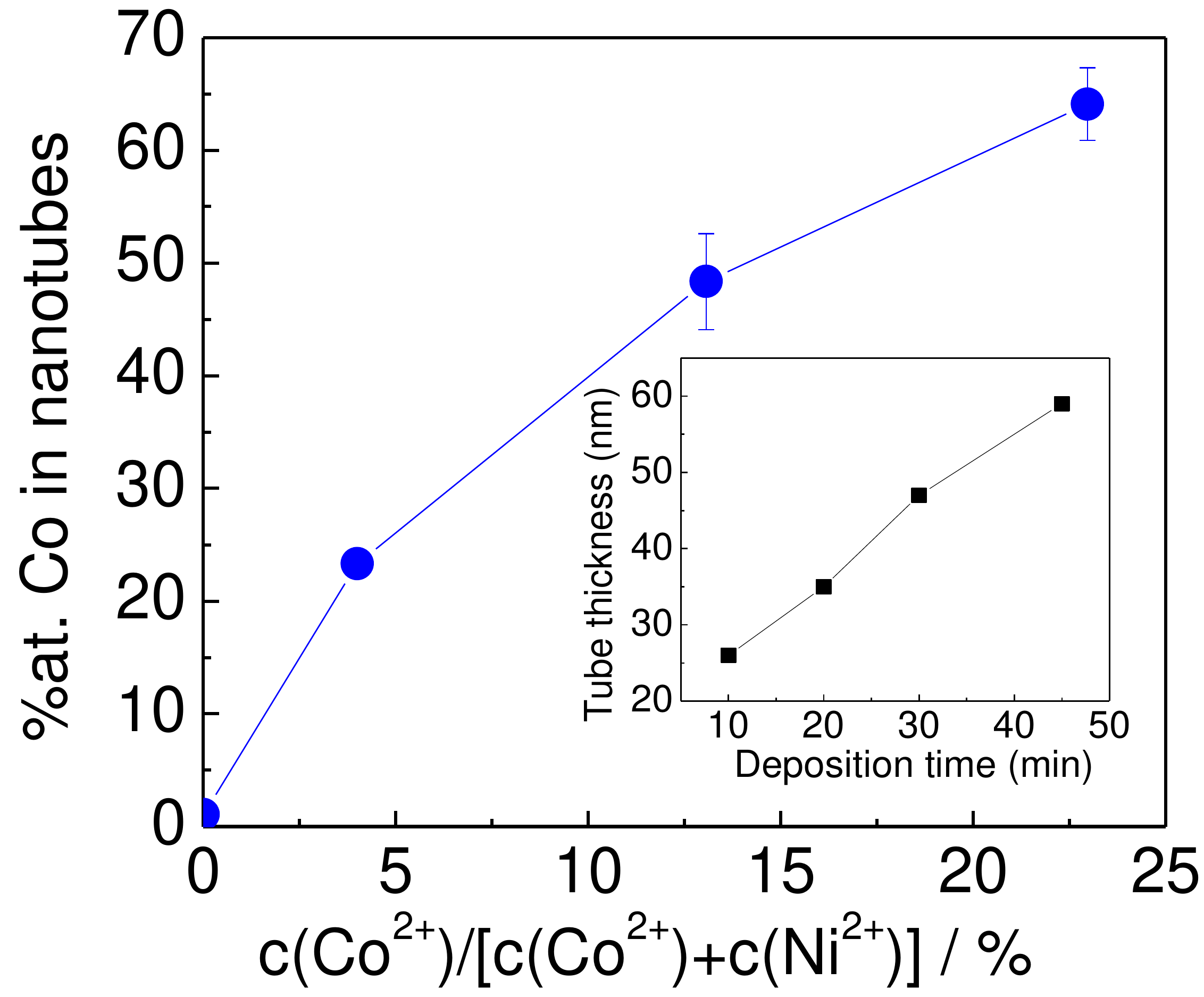}
\caption{\label{fig:Co_content} Co-content of the final NiCoB NTs in relation to the \ch{Co^{2+}} content in the electrolyte, showing the preferential deposition of Co. Inset shows the variation of tube thickness with respect to deposition time for a \ch{(Ni_{0.3}Co_{0.7})B}} deposit.
\end{figure}

In practice, this might cause problems for Ni-rich deposits, as high plating speeds can cause inhomogeneous thickness along the tubes or even lead to a blockage of the pore openings. In order to alleviate this effect, the Ni-rich depositions were conducted at lower temperatures, enabling a more controlled deposition. The inset in Fig.~\ref{fig:Co_content} shows the thickness of the deposit (i.e. the NT wall thickness) in relation to the plating time for a \ch{(Ni_{0.3}Co_{0.7})B} deposit. Although in the time frame observed in our study the wall thickness appears to change linearly with time, it is expected that the plating reaction slows down after a while, due to the ongoing consumption of both metals salts and reducing agent.

The relation between the \ch{Co^{2+}}-content in the electrolyte and the Co-content in the final tubes is given in Fig.~\ref{fig:Co_content}. Due to the electrochemical similarities between \ch{Co^{2+}} and \ch{Ni^{2+}}, one might expect that the Co-content in the deposit either linearly follows the \ch{Co^{2+}}-content in the electrolyte, or that Ni is deposited predominantly due to its higher reduction potential and catalytic activity. However, it can be observed that Co is deposited preferentially, despite being the less noble of the two metals. This anomalous preferential deposition of Co has been observed before, both in electroless plating~\cite{bib-SAI1998, bib-SCH2016c,bib-SAN2004b} as well as electroplating~\cite{bib-WAN2005d,bib-GOM1998}. One possible explanation for this phenomenon is the adsorption of (intermediate) Co-species on the deposited Ni, hindering further Ni-deposition~\cite{bib-GOM1998}.

The use of DMAB as a reducer leads to the incorporation of B into the deposit~\cite{bib-SAI1998}. In fact, the material likely consists of a complex phase mixture of Ni and Co alloys with different Ni and Co borides featuring a nanocrystalline structure, which appears almost amorphous in X-ray diffraction experiments.~\cite{bib-SAI1998, bib-SCH2016c,bib-VIT2012}  Depending on the plating bath composition and reaction parameters such as pH and temperature, deposits with vastly different B-contents can be realized. According to Richardson~\textit{et al.}~\cite{bib-RIC2015c}, the B-content can be tuned in a wide range by adjusting the pH-value of the plating bath, leading up to \SI{45}{\percent}at.~B using a pH~7.5 electrolyte. Compared to our study, however, they use different additives as well as a much higher relative concentration of DMAB in the plating bath, which can lead to an increased B concentration~\cite{bib-SAI1998}. Other studies that utilize similar bath chemistry to our approach, found lower B-contents in the range of \SIrange{12}{30}{\percent}at., depending on the plating parameters~\cite{bib-SAN2004b, bib-SAI1998, bib-OSA2003}. As we based our synthesis on the recent study by Stano~\textit{et al.}~\cite{bib-FRU2018g}, using the same bath composition and reaction parameters, we expected the B-content to be in the range of \SIrange{10}{25}{\percent}at. To get a better understanding of the B content in our samples, X-ray photo-electron spectroscopy (XPS) was performed on a typical Co-rich deposit (see appendix, section~\ref{sec:XPS}).
In the aforementioned study by Stano~\textit{et al.}, the deposit also has been investigated structurally by TEM, highlighting grain sizes in the range of \SI{10}{\nano\meter} separated by \SIrange{1}{2}{\nano\meter} thick transitional regions, presumably rich in lighter elements, such as O and B~\cite{bib-FRU2018g}. Based on the observed dimensions, it can be roughly estimated that these transitional regions make up between \SIrange{30}{60}{\percent} of the total volume of the deposit~\cite{bib-LU1996}, meaning they likely strongly influence the overall electrical and magnetic properties of the material.


\section{X-ray photo-electron spectroscopy (XPS) analysis of electroless \ch{(Ni_x Co_{1-x})B}}
\label{sec:XPS}

\begin{figure}[h!]
\centering
\includegraphics[width=85mm]{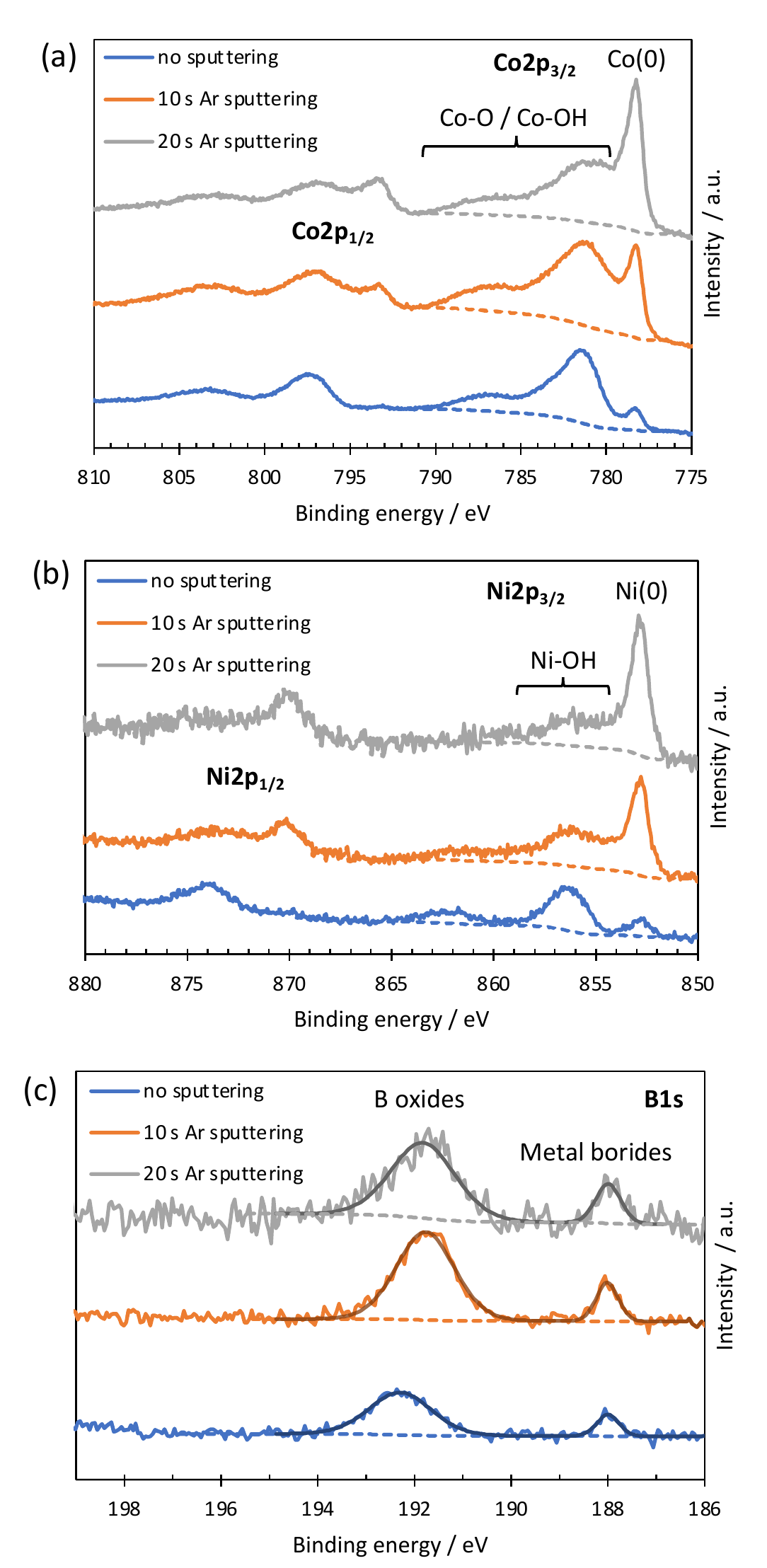}
\caption{\label{fig:XPS} Detailed XPS spectra of the (a) Co2p, (b) Ni2p and (c) B1s regions after 0, 10 and 20~s of Ar sputtering. The data suggest that the deposit consists of a complex phase mixture of superficially oxidized Ni and Co, Ni and Co borides as well as boron oxides. B contents of $\SI{17.60}{\percent}$at., $\SI{19.54}{\percent}$at., and $\SI{19.64}{\percent}$at. can be determined after \SI{0}{\second}, \SI{10}{\second}, and \SI{20}{\second} of Ar sputtering, respectively. The subtracted backgrounds for composition analysis are shown as dashed lines, peak fitting is performed for the B1s line (c), depicted as darker solid lines.}
\end{figure}

XPS was performed on a typical electroless Co-rich deposit, in order to investigate the chemical configuration as well as the B content of the material. Due to the surface sensitivity of the technique, three measurements were conducted with intermittent Ar sputtering for \SI{10}{\second}.

\subsection*{Measurement parameters}

All measurements were performed using a monochromatic X ray source (Al K$\alpha$) with an excitation energy of \SI{1486.6}{\electronvolt} at a \textit{Thermo Fisher Scientific} Escalab 250 spectrometer using a spot size of 650 µm. Pass energies of \SI{10}{\electronvolt} and step sizes of \SI{0.05}{\electronvolt} with a dwell time of \SI{50}{\milli\second} per measurement point were used. Ar sputtering was performed inside the XPS measurement chamber using a \textit{Thermo Fisher Scientific} EX05 ion gun. The acceleration voltage and spot size were set to \SI{3}{\kilo\electronvolt} and $3\times\SI{3}{\milli\meter}$, respectively.
All spectra were calibrated to the Fermi level of silver (\SI{0}{\electronvolt}), the binding energy of the Au4f$_{7/2}$ emission line (\SI{84.0}{\electronvolt}), the Ag3d$_{5/2}$ emission line (\SI{368.26}{\electronvolt}) and the Cu2p$_{3/2}$ emission line (\SI{932.67}{\electronvolt}). Background subtraction and fitting was performed using \textit{CasaXPS} Version 2.3.16Dev52. A Shirley background was applied for all emission lines. For Co2p and Ni2p, the background subtracted spectra were integrated, meaning that no peak fitting was performed. To determine the peak areas, the resulting background corrected spectra were then simply integrated. For fitting the B1s emission lines, GL(30) line shapes were used. To normalize the peak areas of the different elements, the areas were divided by the respective Scofield sensitivity factors, the energy-dependent spectrometer transmission function and KE$^{0.6}$, with KE being the kinetic energy of photo-electrons, to account for the energy-dependent mean free path.

\begin{table}[h!]
   \caption{Concentrations of Co, Ni and B in the investigated Co-rich deposit, as determined by XPS.}
     \centering
     \begin{tabular}{clll}
       \textbf{Ar sputtering / s}  & \textbf{Co / \%at.} & \textbf{Ni / \%at. } & \textbf{B / \%at.}\\
          \toprule
           0    &   74.08   &   8.32    &   17.60 \\
           10   &   71.71   &   8.75    &   19.54 \\
           20   &   71.67   &   8.69    &   19.64 \\
          \bottomrule
     \end{tabular}
     \label{tbl:XPS_composition}
   \end{table}

\subsection*{Results of XPS analysis}

The Co2p and Ni2p lines suggest the presence of metallic and oxidic species of both elements. As shown in Fig.~\ref{fig:XPS}~(a) the Co2p$_{3/2}$ line is divided into multiple peaks. Here, the peak at 778.3 eV is attributed to metallic Co, whereas peaks in the range from 780 to 790 eV suggest the presence of Co oxides. In the case of Ni (Fig.~\ref{fig:XPS}~(b)), a similar behavior can be observed, with a metallic peak at 852.9 eV and oxidic contributions from 855 to 860 eV. The position of the oxidic peaks suggests that the dominant species in this case is likely $\ch{Ni(OH)_2}$~\cite{bib-WAN1998}. The 2p$_{1/2}$ lines of both elements further corroborate the coexistence of metallic and oxidic species. As the ratio between metallic and oxidic contributions shifts towards the former with increasing sputter time, it can be assumed that the metal oxides form superficially after synthesis due to the reaction with atmospheric oxygen and moisture. This agrees with the previously discussed findings from EELS analysis, showing the absence of metal oxides in the bulk material. It is worth noting, however, that the Ar sputtering could also partially contribute to the reduction of both Co and Ni oxides. The B1s line [Fig.~\ref{fig:XPS}~(c)] is separated into two peaks, clearly hinting at the presence of two distinct B species. The peak at higher binding energies (around $\SI{192}{\electronvolt}$) can be attributed to boron oxides (in particular $\ch{B_2 O_3}$), while the peak at around $\SI{188}{\electronvolt}$ indicates the presence of Co and Ni borides~\cite{bib-LEG2002, bib-CHE2005b, bib-SCH2021}. This dichotomy of B species is commonly observed in this type of material, where the B oxides likely are a product of NiB and CoB oxidation~\cite{bib-LEG2002, bib-GLA1994}. Since independent fitting of the metallic and oxidic Ni2p and Co2p peaks is challenging, only a Shirley background subtraction was performed and the resulting spectra were then integrated to determine the total peak areas. The B content amounts to $\SI{17.60}{\percent}$at., $\SI{19.54}{\percent}$at., and $\SI{19.64}{\percent}$at. after \SI{0}{\second}, \SI{10}{\second}, and \SI{20}{\second} of Ar sputtering, respectively (see Table.~\ref{tbl:XPS_composition}). This lies well within the range of B concentrations reported in electroless CoB and NiB alloys fabricated using citrate and DMAB as stabilizer and reducer, respectively~\cite{bib-SAI1998, bib-OSA2003, bib-SAN2004b}.


%

\end{document}